\documentclass[preprint, 12pt, sort&compress]{elsarticle}
\usepackage[utf8]{inputenc}

\date{January 20, 2023}

\usepackage{algorithm}
\usepackage{algorithmic}
\usepackage{amssymb}
\usepackage{bbm}
\usepackage{bm}
\usepackage{cancel}
\usepackage{color}
\usepackage{graphicx}
\usepackage{hyperref}
\usepackage{mathtools}
\usepackage[percent]{overpic}
\usepackage{soul}
\usepackage{subcaption}
\usepackage[symbol]{footmisc}
\usepackage{tikz}
\usepackage{xcolor}

\makeatletter
\def\ps@pprintTitle{%
  \let\@oddhead\@empty
  \let\@evenhead\@empty
  \def\@oddfoot{\reset@font\hfil\thepage\hfil}
  \let\@evenfoot\@oddfoot
}
\makeatother

\hypersetup{unicode=true,bookmarksnumbered=true,bookmarksopen=true,breaklinks=false,colorlinks=false,bookmarksdepth=3}
\numberwithin{equation}{section}
\begin{document}
	
	\begin{frontmatter}
		\title{Controlling auxeticity in curved-beam metamaterials via a deep generative model}
  
		\author{Gerrit Felsch$^1$, Naeim Ghavidelnia$^1$, David Schwarz$^1$, Viacheslav Slesarenko$^{1,*}$ \\[0.6em] \footnotesize $^1$\textit{Cluster of Excellence livMatS @ FIT – Freiburg Center for Interactive Materials and Bioinspired Technologies, University of Freiburg, Georges-Köhler-Allee 105, D-79110 Freiburg, Germany} }

        \begin{abstract}
            Lattice-based mechanical metamaterials are known to exhibit quite a unique mechanical behavior owing to their  rational internal architecture. This includes unusual properties such as a negative Poisson's ratio, which can be easily tuned in reentrant-hexagonal metamaterials by adjusting the angles between beams. However, changing the angles also affects the overall dimensions of the unit cell. We show that by replacing traditional straight beams with curved ones, it is possible to control the Poisson's ratio of reentrant-hexagonal metamaterials without affecting their overall dimensions. While the mechanical properties of these structures can be predicted through finite element simulations or, in some cases, analytically, many applications require to identify architectures with specific target properties. To solve this inverse problem, we introduce a deep learning framework for generating metamaterials with desired properties. By supplying the generative model with a guide structure in addition to the target properties, we are not only able to generate a large number of alternative architectures with the same properties, but also to express preference for a specific shape. Deep learning predictions together with experimental measurements prove that this approach allows us to accurately generate unit cells fitting specific properties for curved-beam metamaterials.
		\end{abstract}
	\end{frontmatter}
    \footnotesize
	*Corresponding Author\\
    E-mail address: \texttt{viacheslav.slesarenko@livmats.uni-freiburg.de}
	\normalsize
 
	\section{Introduction}
 
    The extreme mechanical properties and unique behavior of mechanical metamaterials originate in their involved internal organization \cite{zadpoor_mechanical_2016}. One of the most well-known and well-studied manifestations of unusual behavior is a negative Poisson's ratio in so-called auxetics \cite{kolken_auxetic_2017}. In contrast to conventional materials that shrink laterally while subjected to uniaxial tension, auxetic materials undergo lateral expansion \cite{ren_auxetic_2018}. The key to the auxetic behavior is hidden in the rational internal architecture often consisting of primitive building blocks -- unit cells. While it is possible to combine diverse unit cells within the architecture to achieve specific macroscopic behavior \cite{wilt_accelerating_2020, mirzaali_shape-matching_2018, zhang_programmable_2022}, traditional auxetic design is usually based on a single unit cell that repeats itself in two or three dimensions \cite{zheng_ultralight_2014}. In general, a wide variety of unit cell designs (reentrant \cite{korner_systematic_2015}, chiral \cite{reinbold_rise_2019, wu_mechanical_2019}, etc.) in 2D and 3D settings that transform uniaxial elongation to lateral expansion were proposed. Moreover, in 3D settings, unit cells can couple compression and twist deformations \cite{frenzel_three-dimensional_2017, fernandezcorbaton_new_2019}. It is important to notice, that while  elastic constants such as the Poisson's ratio were originally introduced for infinitesimal deformations in isotropic (or orthotropic) materials, a conceptually similar value connecting traversal and lateral deformations has been actively employed to describe mechanical behavior for large deformations \cite{ling_experimentally_2020} or for anisotropic metamaterials \cite{mirzaali_multi-material_2018}. Assuming applicability of the Poisson's ratio for large deformations, it was shown that it is possible to facilitate transition from positive to negative Poisson's ratio via elastic buckling \cite{li_auxetic_2018} or external non-mechanical stimuli \cite{skarsetz_programmable_2022, he_electrospun_2021} in mechanical metamaterials.   
    
	For traditional mechanical metamaterials \cite{korner_systematic_2015,reinbold_rise_2019,wu_mechanical_2019}, the geometry of the unit cell univocally defines the overall mechanical behavior. Therefore, it is enough to perform corresponding analysis only for a single unit cell assuming periodic boundary conditions \cite{mizzi_implementation_2021, ryvkin_fault-tolerant_2020}. For the classical reentrant-hexagonal unit cell design shown in Fig. \ref{fig:Hexagon_Reentrant}, analytical solutions for the stiffness matrix were obtained under the assumptions of Euler-Bernoulli or Timoshenko beam formulations \cite{ghavidelnia_idealized_2021, hedayati_improving_2021, mukherjee_general_2021}, or via other homogenization methods \cite{berinskii_-plane_2018, berinskii_elastic_2020}. In particular, it is established that the metamaterial demonstrates auxetic behavior only if the corresponding angle $\theta$ is negative (Fig. \ref{fig:Hexagon_Reentrant}b); otherwise, the metamaterial possesses a positive Poisson's ratio (Fig. \ref{fig:Hexagon_Reentrant}a). Harnessing this dependency, metamaterials capable of reversibly switching their auxeticity were realized with the help of stimuli-responsive materials embedded into the architecture \cite{skarsetz_programmable_2022,wang_three-dimensional_2020}.
    
    While the reentrant-hexagonal architecture provides an easy way for programming its Poisson's ratio via alteration of the angle, such geometrical change affects the overall size of the metamaterial. More specifically, it is impossible to obtain two distinct reentrant-hexagonal metamaterials with different Poisson's ratios that simultaneously share the common overall dimensions and number of unit cells. In this manuscript, we will reveal how minor modification of the reentrant-hexagonal architecture might enable a very wide range of admissible Poisson's ratios while keeping the dimensions of the unit cell constant. Instead of the classical unit cells shown in Fig. \ref{fig:Hexagon_Reentrant}, we will consider unit cells constructed with the help of curved beams. 

    The versatility of the design for non-straight beams and the resulting intriguing interplay between their bending and lateral stiffnesses enable the enriched potential for programming the behavior of mechanical metamaterials through a rational selection of the geometry of the unit cell \cite{fu_design_2021, mukherjee_-plane_2022}. For example, incorporating Bezier curves in designs has been shown to widen the design space for chiral metamaterials \cite{alvarez-trejo_bezier_2021}. While the benefits of curved beams are quite obvious, their employment in the structure makes the analytical solution for the mechanical behavior of metamaterials much harder \cite{mukherjee_-plane_2022, harkati_out--plane_2021}. Therefore, numerical methods, such as finite element analysis (FEA), are widely used to find corresponding mechanical constants. However, besides a forward problem to obtain the properties for a specific geometry, from the application point of view, the inverse question of searching for the geometry that will provide the requested mechanical response is even more important. Since multiple curves might generate unit cells with the same properties, the inverse problem becomes ill-posed and requires more advanced solution methods. 

    The recent development of generative models in machine learning (ML) provides ideas on how to tackle this issue, with variational autoencoders (VAEs) \cite{kingma_auto_2014} and generative adversarial networks (GANs) \cite{goodfellow_generative_2020} having shown promising results for the generation of mechanical metamaterials \cite{wang_deep_2020,wang_ih-gan_2022}. These models learn a transformation to a latent space, where each point corresponds to a valid design. When applied to metamaterials, the dimensions of this space are usually connected to geometric features, which easily allows to interpolate between different designs \cite{wang_deep_2020}.   However, learning a mapping, where some dimensions reflect mechanical properties instead of geometric features, is computationally expensive, and learning algorithms can be unstable. As an alternative, it has been shown that through the smart design of neural networks and rational selection of the loss functions, it is possible to find a suitable mapping between properties and structure without learning it from a scratch \cite{kumar_inverse_2020, bastek_inverting_2022}.
    In this manuscript, we propose an approach that enables us to efficiently predict the mechanical response of reentrant-hexagonal metamaterials based on curved Bezier beams and, more importantly, efficiently search for a wide variety of geometries that will suffice the specific requirements.
	
	\section{Unit cell design}
	
    \subsection{Classical hexagonal and reentrant lattices}
	
\begin{figure}[ht]
		\centering
        \begin{minipage}[t]{0.49\textwidth}  
		\hspace{0.05em} \textbf{(a)} \\  
		\begin{minipage}[t]{\textwidth}
            \centering
			\includegraphics[height=0.19\textheight]{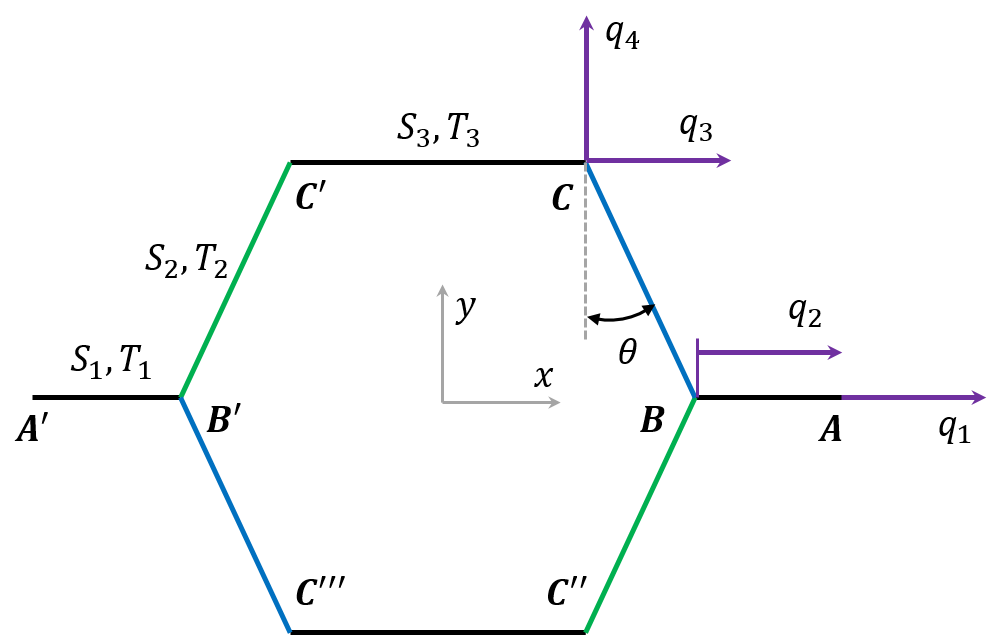}
		\end{minipage}
         \end{minipage}
         \begin{minipage}[t]{0.49\textwidth}  
		\hspace{0.05em} \textbf{(b)} \\ 
		\begin{minipage}[t]{\textwidth}
            \centering
            \vspace{1.15em}
			\includegraphics[height=0.145\textheight]{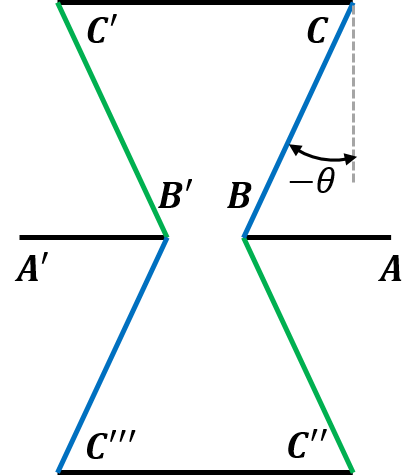}
		\end{minipage}
        \end{minipage}
        \begin{minipage}[t]{0.98\textwidth}  
        \vspace{2.5em} 
		\hspace{0.05em} \textbf{(c)} \\ 
            \begin{minipage}[t]{1\textwidth}
         	      \centering
			      \includegraphics[height=0.14\textheight]{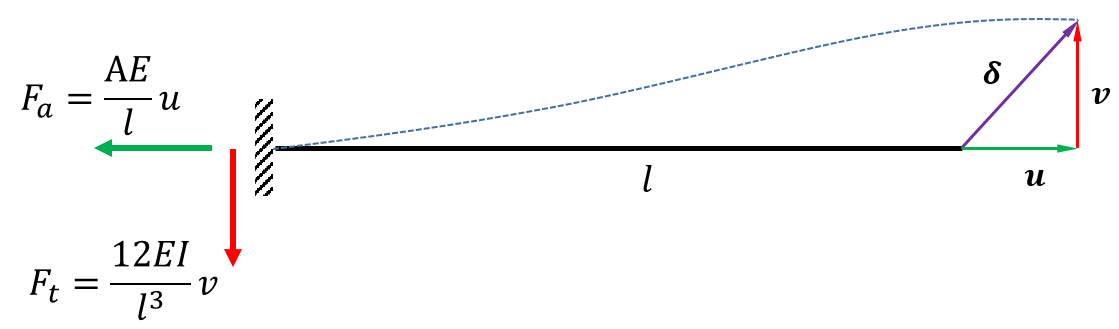} 
            \end{minipage}
		\end{minipage} 
		\caption{Classical hexagonal unit cell with its degrees of freedom \textbf{(a)}. Classical reentrant unit cell \textbf{(b)}. Axial and lateral effective stiffness of the general beam \textbf{(c)}}
		
		\label{fig:Hexagon_Reentrant}
	\end{figure}

	A honeycomb structure is a well-known two-dimensional cellular lattice based on the six-sided hexagonal polygon unit cells with the identical strut length and internal angles. Any change in the lengths and angles of the honeycomb structure can lead to a new structure with different mechanical properties. Fig.~\ref{fig:Hexagon_Reentrant}a and Fig.~\ref{fig:Hexagon_Reentrant}b illustrate the resultant classical hexagonal and reentrant unit cells by altering the angle of inclined struts (BC) from positive to negative values of $\theta$, which yield a wide range of different structures with positive and negative Poisson's ratios. The analytical solution for a hexagonal/reentrant structure with equal cross section of the beams, limited height and limited width of the unit cell, has been derived previously by Hedayati et al. \cite{hedayat_analytical_2020}. In this section, a general solution for hexagonal/reentrant structures that consist of various independent lengths and cross sections for beams is derived. Due to the intrinsic geometrical symmetries of the hexagonal lattice, the behavior of the corresponding unit cell can be described by four individual degrees of the freedom (Fig.~\ref{fig:Hexagon_Reentrant}a). 
	
    By considering small deflections and linear material behavior, the overall deformation of the unit cell can be considered as the superposition of separate deformations caused by applying individual loads at each degree of freedom. Therefore, the superposition principle could be implemented for obtaining the system of equations for the unit cell. The system of equilibrium equations for the structure could be written as:
    
    \begin{align}
    	\left\{
            \begin{array}{c}
             Q_{1} \\
             Q_{2} \\
             Q_{3} \\
             Q_{4} \\
            \end{array}
        \right\}=\left(
            \begin{array}{cccc}
             k_{11} & k_{12} & k_{13} & k_{14} \\
             k_{21} & k_{22} & k_{23} & k_{24} \\
             k_{31} & k_{32} & k_{33} & k_{34} \\
             k_{41} & k_{42} & k_{43} & k_{44} \\
            \end{array}
        \right)\left\{
            \begin{array}{c}
             q_{1} \\
             q_{2} \\
             q_{3} \\
             q_{4} \\
            \end{array}
        \right\}\label{eqsystem}
    \end{align}    
	 
    In this equation, $\{Q\}$ is the force vector which is composed of the external forces acting on the unit cell, $[K]$ is the stiffness matrix of the system and $\{q\}$ is the displacement vector corresponding to each degree of freedom. To calculate the elements of the stiffness matrix of the unit cell based on the superposition method, unit displacements are applied individually to each degree of freedom  ($q_{i}= 1, i=1,2,3,4$) by constraining all the other degrees of freedom ($q_{j}= 0, j\not=i$). Due to the symmetrical planes of the hexagonal unit cells, there is no rotational degree of freedom at the joints, which means that the struts of the unit cells are either under elongation (or contraction) or bending (without rotation) at the joints. Therefore, the deformation of an arbitrary strut within the unit cell can be illustrated as shown in Fig.~\ref{fig:Hexagon_Reentrant}c. The general deformation $\delta$ at the free end of the beam, is decomposed into an axial 
    ($u$) and a lateral ($v$) deformation, which yields the corresponding reaction forces ($F_{a}$, $F_{t}$) at the fixed end of the beam. Based on the Euler-Bernoulli beam theory, the axial and lateral stiffness of the beam can be derived as $S_i=\frac{ A_{i}E}{l_i}$ and $T_i=\frac{12EI_{i}}{l_i^3}$ respectively.

    By solving the force equilibrium equations for each beam of the unit cell according to the reaction forces resulting from individual degrees of freedom, the stiffness matrix of the unit cell is extracted as follows:

	\begin{align}
    	\resizebox{1\hsize}{!}{%
    	$K=\left(
            \begin{array}{cccc}
             2 S_1 & -2 S_1 & 0 & 0 \\
             -2 S_1 & 4 S_2 \sin ^2(\theta )+2 S_1+4 T_2 \cos ^2(\theta ) & -4 S_2 \sin ^2(\theta )-4 T_2 \cos ^2(\theta ) & \sin (\theta ) \cos (\theta ) \left(4 S_2-4 T_2\right) \\
             0 & -4 S_2 \sin ^2(\theta )-4 T_2 \cos ^2(\theta ) & 4 S_2 \sin ^2(\theta )+4 S_3+4 T_2 \cos ^2(\theta ) & \sin (\theta ) \cos (\theta ) \left(4 T_2-4 S_2\right) \\
             0 & \sin (\theta ) \cos (\theta ) \left(4 S_2-4 T_2\right) & \sin (\theta ) \cos (\theta ) \left(4 T_2-4 S_2\right) & 4 S_2 \cos ^2(\theta )+4 T_2 \sin ^2(\theta ) \\
            \end{array}
        \right)
        $}
	\end{align}

    After deriving the  stiffness matrix of the unit cell, the system of equations (Eq. \ref{eqsystem}) can be solved for different loading conditions. To obtain the mechanical properties of the hexagonal structure in $x$ and $y$ directions, the corresponding loading conditions are applied to the unit cell and the system of equations is solved to calculate the resultant deformations at the defined degrees of freedom. The loading conditions of $Q_{1}= 1$ and $Q_{4}= 1$ yield the different results for $q_{i}$ that are used for calculating the mechanical properties of the structure such as Poisson's ratio. As a general definition, the equations for Poisson's ratio of the hexagonal/reentrant unit cell in $x$ and $y$ directions can be calculated by using $\nu_{xy}=-\frac{\epsilon _y}{\epsilon _x}$ and $\nu_{yx}=-\frac{\epsilon _x}{\epsilon _y}$ formulas. The strains in  $x$ and $y$ directions ($\epsilon _y$ and $\epsilon _y$) are calculated by using the results of $q_{1}$ and $q_{4}$ for their corresponding loading conditions ($Q_{1}= 1$ or $Q_{4}= 1$). Hence the final relationships for Poisson's ratios of the unit cell will be as follows:
	
	\begin{align}
        \nu_{yx}=-\frac{\sin (\theta ) \left(T_2-S_2\right) \left(2 l_2 \sin (\theta )+2 l_1+l_3\right)}{2 l_2 S_2 T_2 \left(\frac{\sin ^2(\theta )}{S_2}+\frac{2}{S_1}+\frac{1}{S_3}+\frac{\cos ^2(\theta )}{T_2}\right)}
	\end{align}
	
	\begin{align}
        \nu_{xy}=-\frac{2 l_2 \sin (\theta ) \cos ^2(\theta ) \left(T_2-S_2\right)}{S_2 T_2 \left(2 l_2 \sin (\theta )+2 l_1+l_3\right) \left(\frac{\cos ^2(\theta )}{S_2}+\frac{\sin ^2(\theta )}{T_2}\right)} \label{eq.v21}
	\end{align}

    The aforementioned results are used to calculate the Poisson's ratio of hexagonal structures by using positive values of $\theta$, or reentrant structures by substitution of negative values of $\theta$. The Poisson's ratio in the two directions clearly depends on both the axial and lateral stiffness of the horizontal and inclined beams. Varying these stiffnesses by substituting the classical straight beams with curved beams, a huge design variability can be achieved.   
	
	\subsection{Bezier curve theory}

 	\begin{figure}[t]
		\centering
        \begin{subfigure}[b]{0.6\textwidth}
            \def\svgwidth{\linewidth}  
            \includegraphics[width=\textwidth]{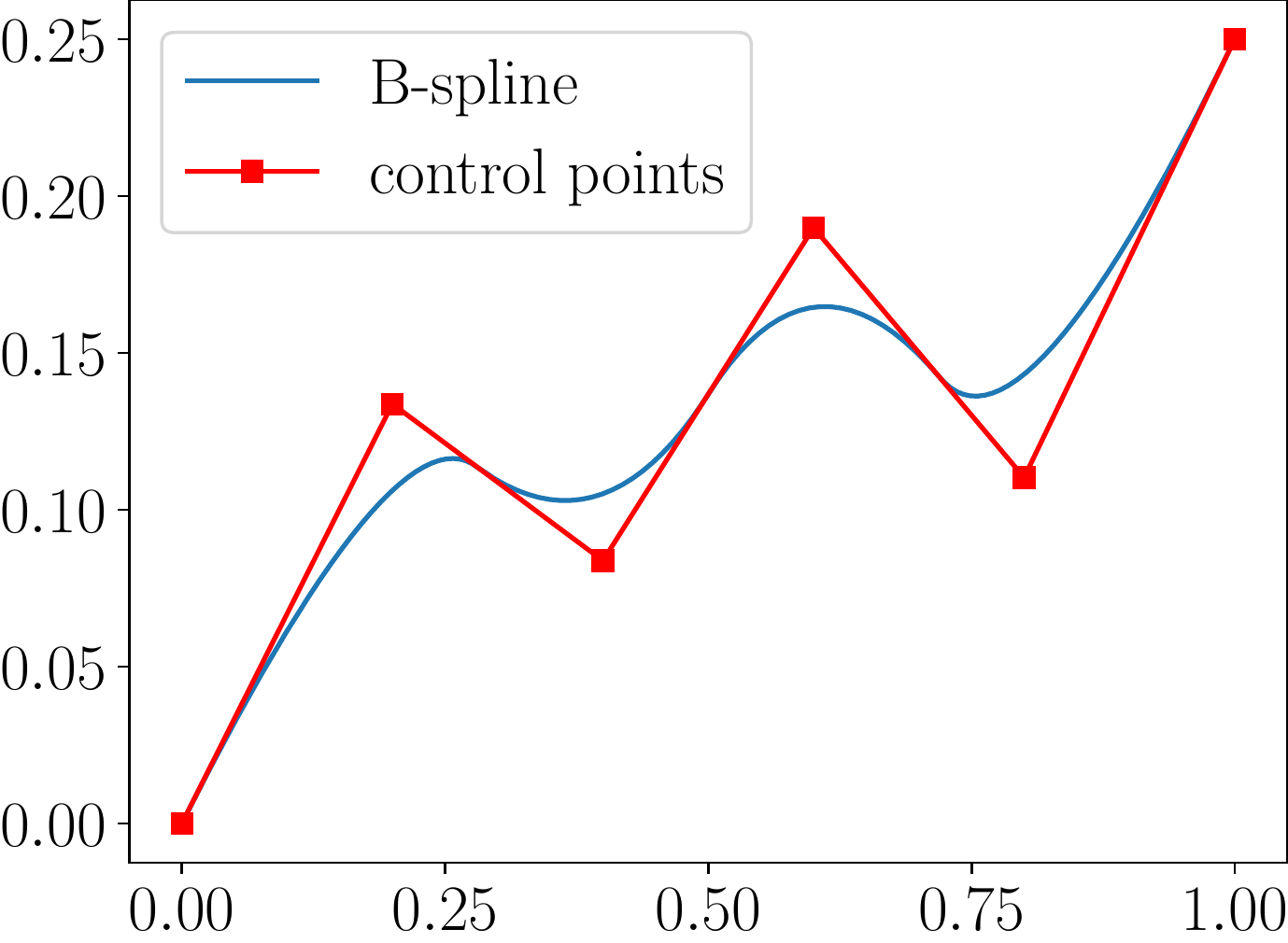}
        \end{subfigure}  
		\caption{A quadratic ($p=2$) B-spline build from the six control points ($c_i, i=1,..,6$)  and the knot vector $k = (0, 0, 0, 0.2, 0.5, 0.8, 1, 1, 1)$}
		\label{fig:b_spline}
	\end{figure} 

    To replace straight beams with complex curved elements, a suitable representation of these elements is needed. B-splines are able to provide smooth parametric descriptions of such elements. By definition, these B-spline curves $B(t)$ are constructed from a combination of so-called control points and several basis functions of order $p$ \cite{hughes_isogeometric_2005}. One such B-spine and the corresponding control points  $c_i \in \mathbb{R}^d$ are shown in Fig. \ref{fig:b_spline}. Each point $t \in [0,1]$ along the curve is given by a weighted sum of the control points, where the weights are defined by the basis functions $N_{i,p}:  [0,1] \rightarrow [0,1] $ as
	
	\begin{align}
	    B(t) = \sum_{i=1}^{n} N_{i,p}(t) c_i \label{eq:b-spline}
	\end{align}

	This method assures that some parts of the curve are more dependent on certain control points than others, which gets more evident when looking at the way the basis functions themselves are constructed via recursion.
    First, the curve is partitioned into smaller intervals, divided by knots $k_i, ~ i=1,\dots,n+p+1, ~ k_i \leq  k_{i+1}$. Each zero-order basis ($p=0$) function is defined as  
	
	\begin{align}
	    N_{i,0}(t) = 
        \begin{cases} 	
            1 & \text{if} ~ k_i < t < k_{i+1}   \\
		      0 & \text{otherwise} 
	    \end{cases} 
	\end{align}
 
	As one can see, the basis functions of order 0 are not smooth. 
	The recursive step then performs a smoothing operation on each base function using one neighboring function, giving continuous curves
	
	\begin{align}
    	N_{i,p}(t) = \frac{t_{}-k_{i}}{k_{i+p}-k_{i}} N_{i,p-1}(t) + \frac{k_{i+p+1}-t_{}}{k_{i+p+1}-k_{i+p}} N_{i+1,p-1}(t)
	\end{align}

    As can be seen, the number of control points contributing to each point of the curve increases with the order. Fig. \ref{fig:b_spline} shows a quadratic spline ($p=2$) where this number is usually three. However, the start and end point of the curve each depend only on a single control point. For this to occur, it is necessary to have $p+1$ knots with $k_i = k_0$ and $k_i=k_{n+p+1}$ respectively. In this case Eq. \ref{eq:b-spline} yields $B(0)=c_1$ and $B(1)=c_n$. To generate the B-splines for the unit cells we used the Python-package Splipy \cite{johannessen_splipy_2020}.
    
	\subsection{Generated metamaterials}
	
\begin{figure}
		\centering
        \small
        \begin{subfigure}[b]{0.9\textwidth}
            \def\svgwidth{\linewidth}  
            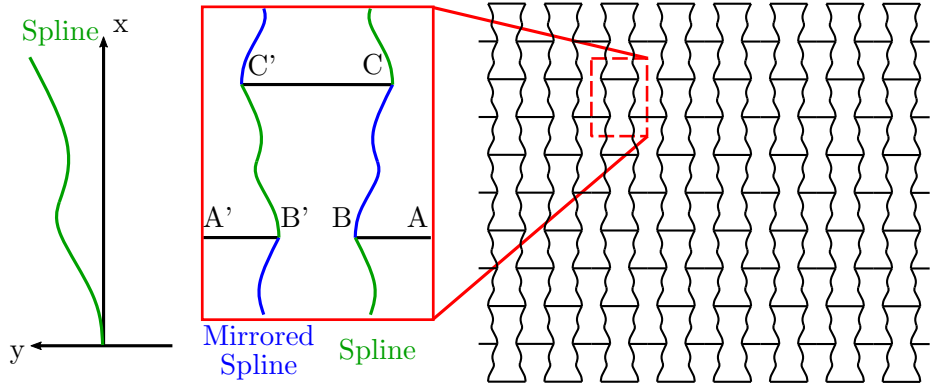
        \end{subfigure}
		\caption{A  B-spline, the corresponding representative unit cell and a resulting system comprised of 8 × 5 representative unit cells.}
		\label{fig:unit_cell}
	\end{figure}

\begin{figure}[t]
		\centering
		\begin{minipage}[t]{0.42\textwidth}  
    		\hspace{0.05em} \textbf{(a)} \\  
    		\begin{minipage}[t]{0.32\textwidth}
                \begin{center}
    			\includegraphics[width=\textwidth]{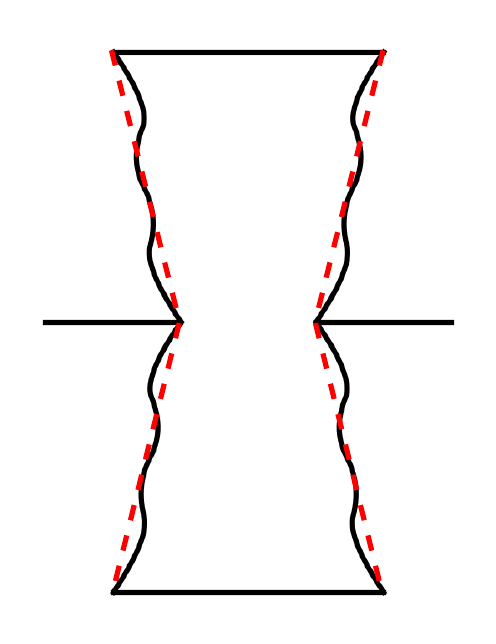}
    		      \end{center}
            \end{minipage}   
    		\begin{minipage}[t]{0.32\textwidth}
                \begin{center}
    			\includegraphics[width=\textwidth]{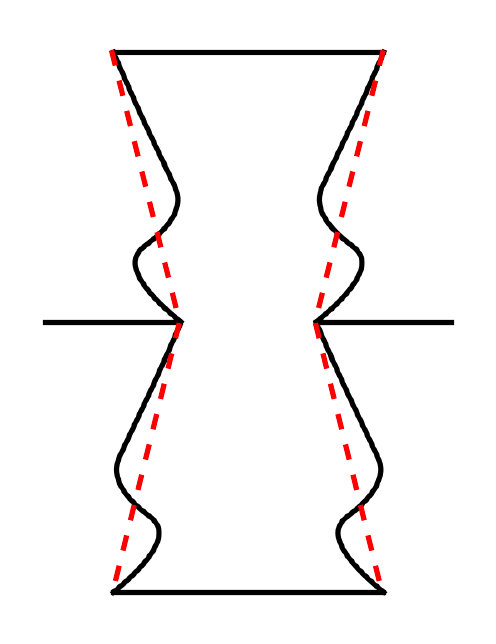}
       		    \end{center}
    		\end{minipage}
    		\begin{minipage}[t]{0.32\textwidth}
                \begin{center}   
    			\includegraphics[width=\textwidth]{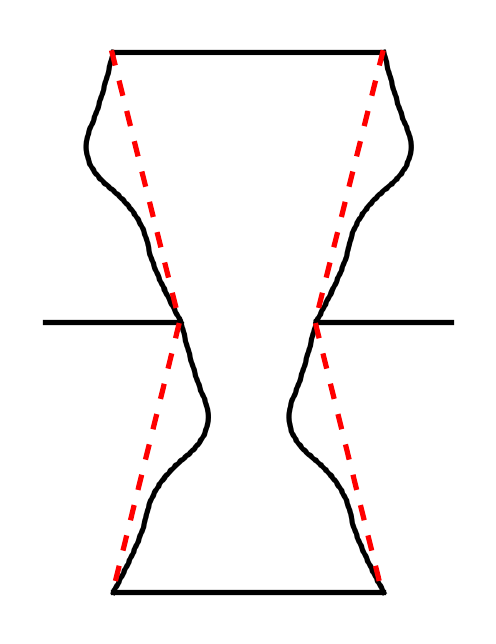}
       		    \end{center}   
    		\end{minipage}  
		\end{minipage}   
        \vline
		\begin{minipage}[t]{0.545\textwidth}  
    		\hspace{0.05em} \textbf{(b)} \\  
    		\begin{minipage}[t]{0.38\textwidth}
                \begin{center}
    			\includegraphics[width=\textwidth]{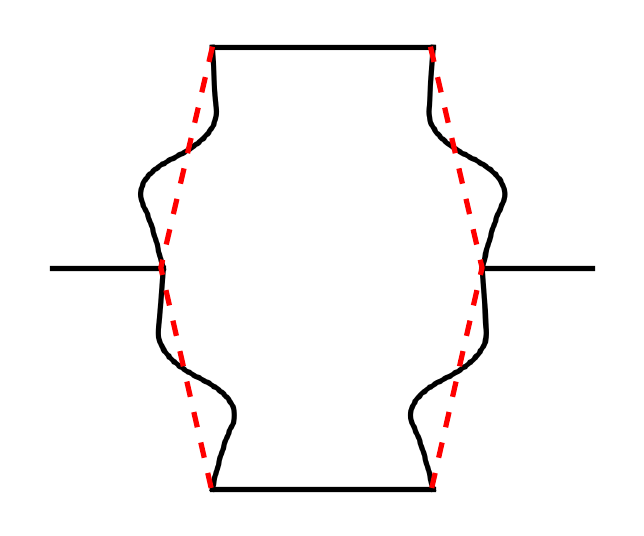}
    		      \end{center}
            \end{minipage}   
    		\begin{minipage}[t]{0.32\textwidth}
                \begin{center}
    			\includegraphics[width=\textwidth]{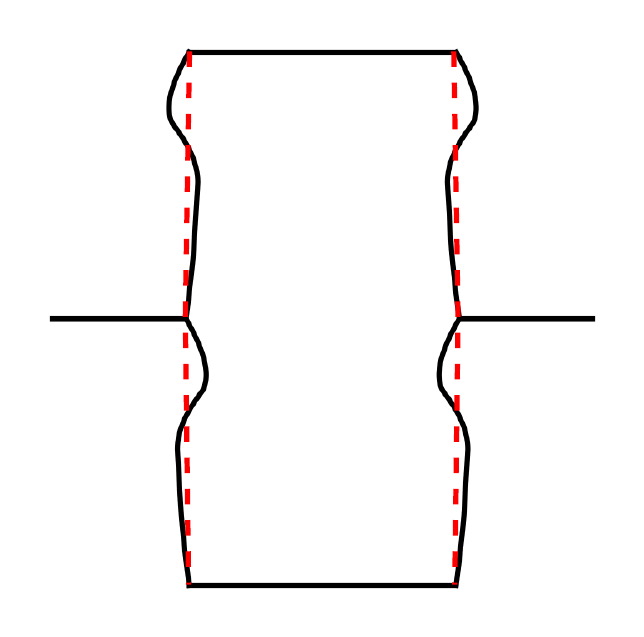}
       		    \end{center}
    		\end{minipage}
    		\begin{minipage}[t]{0.255\textwidth}
                \begin{center}   
    			\includegraphics[width=\textwidth]{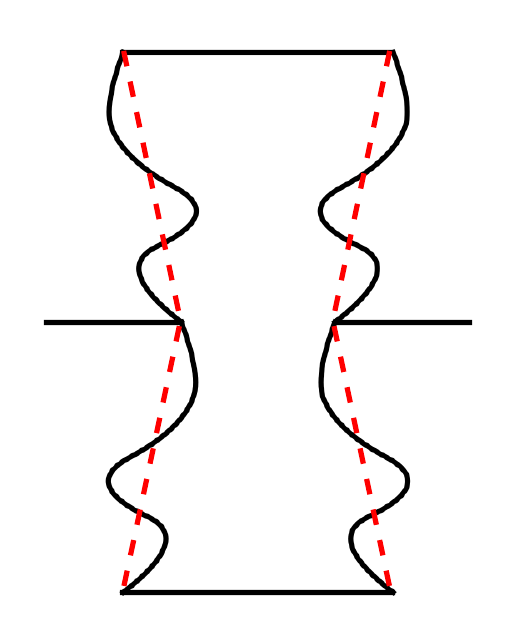}
       		    \end{center}   
		      \end{minipage}  
		\end{minipage}    
		\caption{Unit cell examples from \textit{Set 1} \textbf{(a)} and \textit{Set 2} \textbf{(b)}. While the dimensions of the unit cells from \textit{Set 1} are constant, cells from \textit{Set 2} can resemble both reentrant or hexagonal structures.}
		\label{fig:unit_cell_types}
	\end{figure}

	The unit cells of the metamaterials considered here are inspired by hexagonal and reentrant honeycomb structures. As can be seen in Fig. \ref{fig:unit_cell} the horizontal struts are connected by a curved quadratic B-spline or its mirrored equivalent instead of straight beams. Due to the periodicity, there are several ways to pick a unit cell for the same lattice; the unit cell shown in Fig.  \ref{fig:unit_cell} is just one of them. Like the one shown in Fig. \ref{fig:b_spline}, the B-spline connecting the horizontal struts is given by six control points. As the unit cell is univocally defined by the B-spline, these points can be further used to represent the whole unit cell. For simplification, control points were placed equidistantly along the x-axis, leaving only the coordinates along the y-axis as design variables. Additionally, the length of the horizontal struts ($AB$ in Fig. \ref{fig:unit_cell}) was kept the same for all unit cells, leaving them only dependent on the relative position of the control points. The first control point was always selected as (0,0). Therefore, each structure is given by a set of five independent parameters - the vertical values $c_{i,y}$ of the control points $c_{i}, ~ i = 2,\dots,6$ (see Fig.~\ref{fig:b_spline}). Using these parameters as design variables, two different datasets were created: 	
 
	\begin{itemize}
		\item \textit{Set 1} consists of unit cells based on B-splines with a fixed end point, resulting in unit cells corresponding to variations of a single reentrant unit cell design with $B'B = 0.5 ~ C'C$. For this dataset, the horizontal struts always have the same relative position, leaving the variation of the B-splines as the only source of properties differences. 
		\item \textit{Set 2} consists of unit cells based on B-splines with free end points, resulting in a greater diversity of unit cells. As can be seen in Fig. \ref{fig:unit_cell_types}, varying the y-coordinate of the end of the unit cell can change its underlying geometry from more reentrant like to hexagonal (see the dashed red lines in Fig. \ref{fig:unit_cell_types}b). 
	\end{itemize}

\begin{figure}[t]
		\begin{minipage}[t]{0.49\textwidth}		
            ~ \textbf{(a)} \\
    		\includegraphics[width=\textwidth]{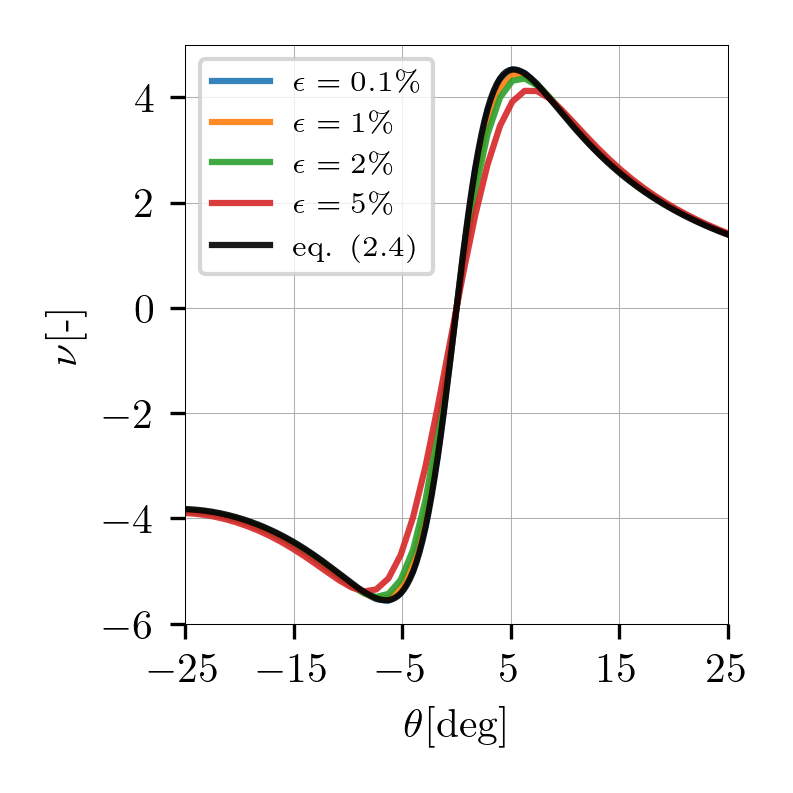}
		\end{minipage}   
		\begin{minipage}[t]{0.49\textwidth}
            ~ \textbf{(b)} \\
    		\includegraphics[width=\textwidth]{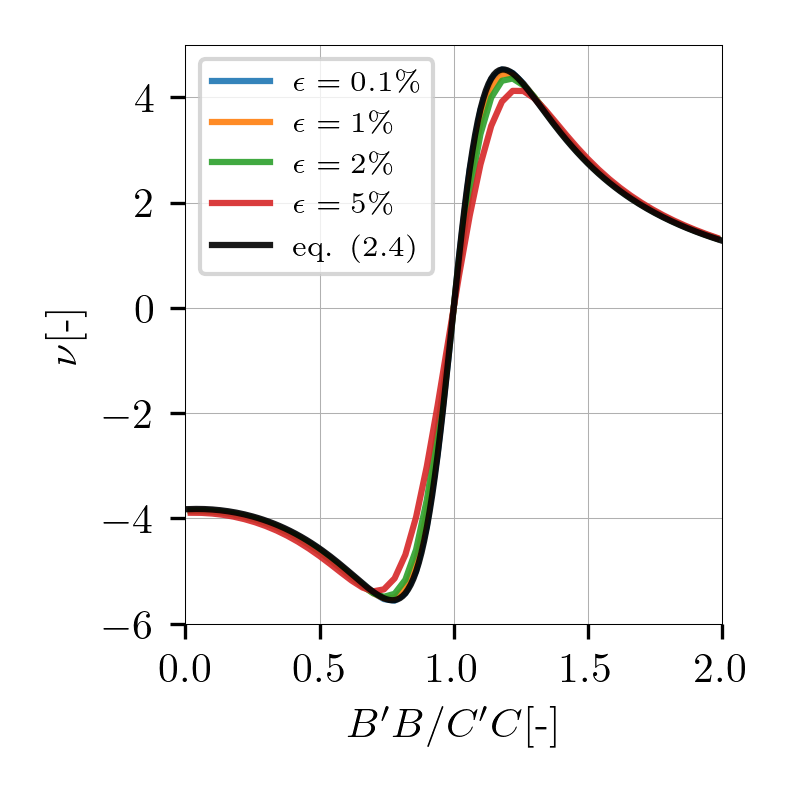}
		\end{minipage}   
		\caption{The Poisson's ratios ($\nu$) of different unit cells with straight beams (as shown in Fig. \ref{fig:Hexagon_Reentrant}a-b)   depending on angle $\theta$ \textbf{(a)} and ratio between $B'B$ and $C'C$ \textbf{(b)} respectively. The black curve corresponds to Eq. \ref{eq.v21}, while the others were obtained through simulations in COMSOL with varying applied strain ($\epsilon$). Note, that simulations match the theory well, especially for small strain.}
		\label{fig:theory}
	\end{figure}

    To obtain the Poisson's ratio for the structures in both datasets, simulations  were performed in the finite element (FE) package COMSOL~5.4a via the LiveLink interface. For these simulations, the B-splines were discretized into 150 beam elements. Periodicity was prescribed by matching the displacements and angles on the boundaries of the unit cell. The Poisson's ratio was found as the ratio between lateral expansion vs applied tensile strain ($-\varepsilon_{yy}/\varepsilon_{xx}$). Note, that the considered metamaterials are not isotropic, therefore the value of this Poisson's ratio is not bounded by 0.5 as in case of isotropic materials. To verify the simulation approach, results for unit cells with straight beams were compared to theoretical values of $v_{xy}$ derived by Eq. \ref{eq.v21}. It was found that simulations match the theory well for small deformations, as shown in Fig. \ref{fig:theory}.

	\section{ML implementation for forward (Structure-Property) problem}
	
\begin{figure}[t]
        \begin{minipage}[t]{0.49\textwidth}	
            \hspace{0.01em} \textbf{(a)} \\[0em]
            \def\svgwidth{\linewidth}
            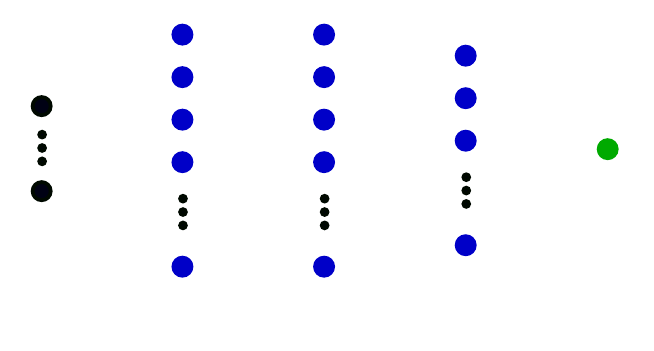     
        \end{minipage}
        \vline      
        \begin{minipage}[t]{0.49\textwidth}	
            ~ \textbf{(b)} \\[0em]  
            \def\svgwidth{\linewidth}
            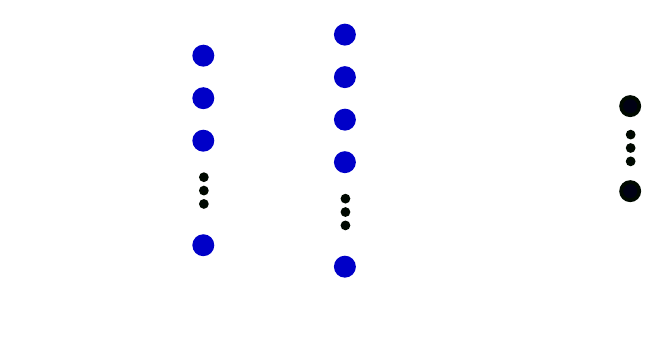    
        \end{minipage}        
        \caption{The forward NN model $\mathcal{F}$ \textbf{(a)} and the inverse NN model $\mathcal{G}$ \textbf{(b)}. The different layers are color coded. Poisson's ratios are shown in green, B-splines in black and hidden layers in blue. The hidden layers consist of fully connected (FC), layer normalization (LN) layers in combination with LeakyReLU (LU) and Sigmoid (Sig) activation functions.}
        \label{fig:fwd_inv_model}    
	\end{figure}

	The Neural Network (NN) model $\mathcal{F}$ depicted in Fig. \ref{fig:fwd_inv_model}a was used to solve the forward problem. This model predicts the Poisson's ratio of a structure upon receiving the corresponding B-spline in form of its five constitutive parameters as input. The network consists of four fully connected layers, including the output layer, where each hidden layer was followed by a ReLU activation function. To learn how to predict the Poisson's ratio, the network was trained using the Adam optimizer \cite{kingma_adam_2015} and a Mean Squared Error objective function. Layer normalization \cite{ba_layer_2016} was used to reduce training time. For both datasets the Poisson's ratios and the B-spline parameters were rescaled to the range [0,1]. On each dataset a separate network with the same architecture was trained.

 	\begin{figure}[t]
		\begin{minipage}[t]{0.49\textwidth}		
            ~ \textbf{(a)} \\
    		\includegraphics[width=\textwidth]{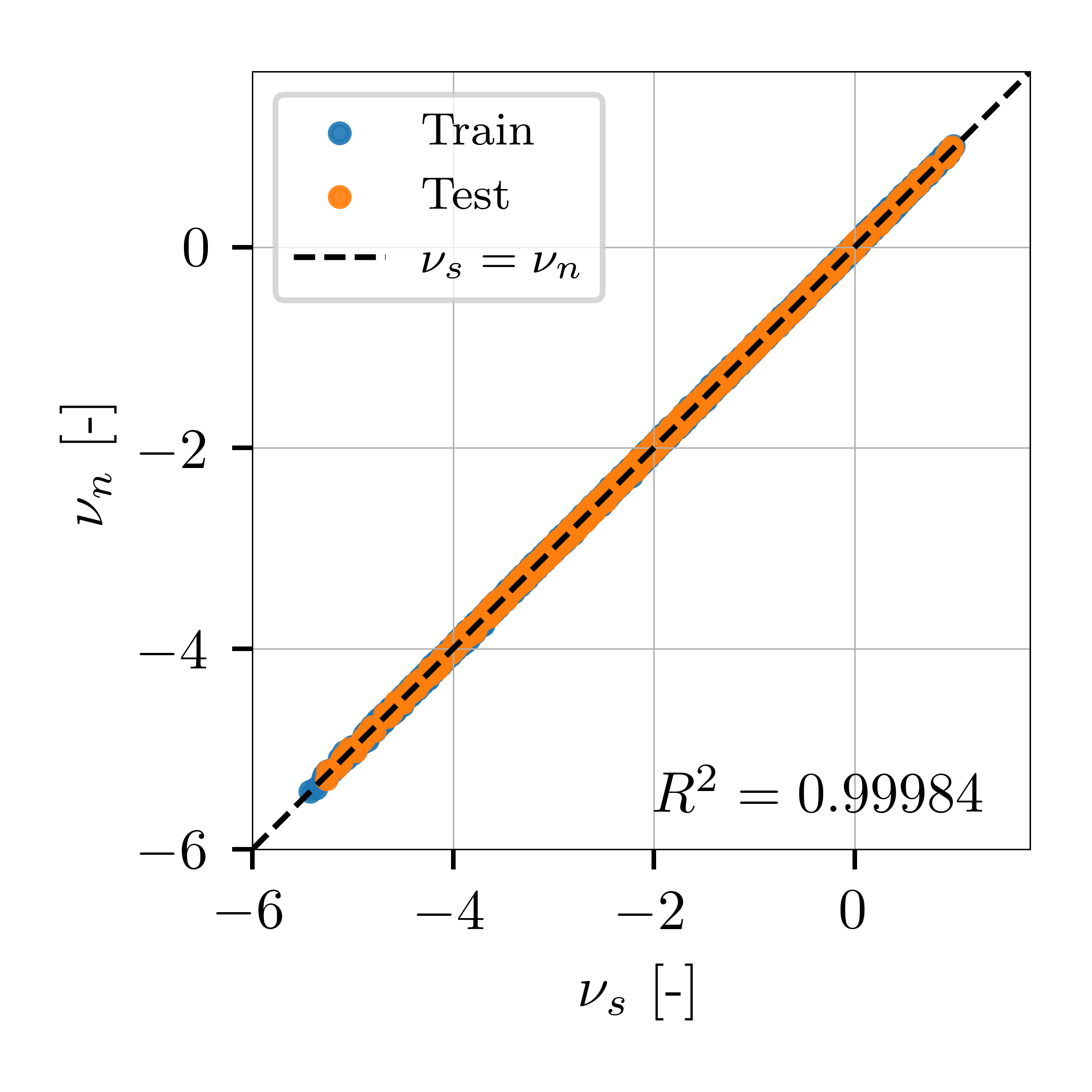}
		\end{minipage}   
		\begin{minipage}[t]{0.49\textwidth}
            ~ \textbf{(b)} \\
    		\includegraphics[width=\textwidth]{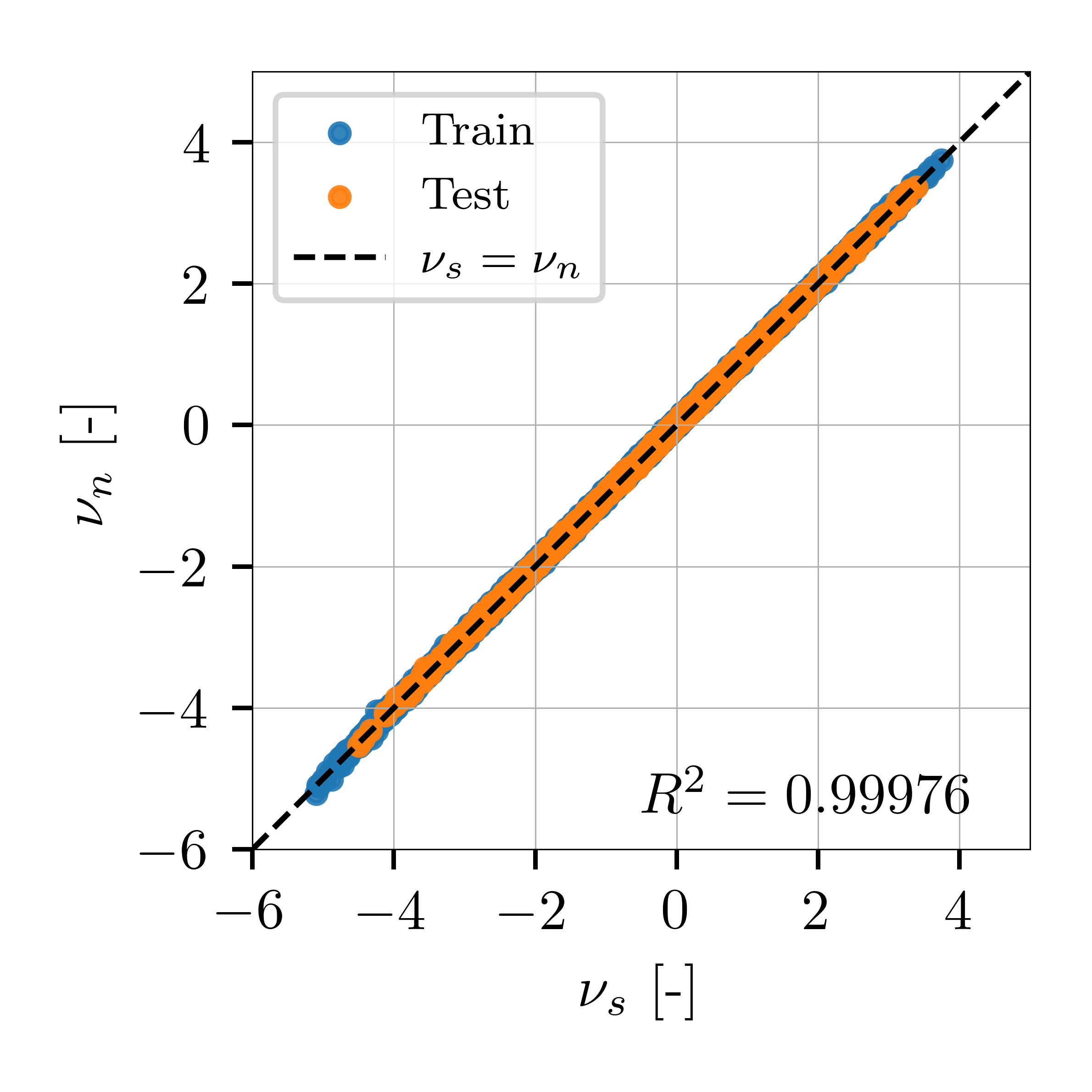}
		\end{minipage}   
		\caption{Comparison between the Poisson's ratios obtained by FEA ($\nu_{s}$) and the forward NN model $\mathcal{F}$ ($\nu_{n}$) for training and testing data from \textit{Set 1} \textbf{(a)} and \textit{Set 2} \textbf{(b)}.}
		\label{fig:forward}
	\end{figure}	
	
	\subsection{Set 1}

    Of the $\text{25,000}$ examples in \textit{Set 1}, $90\%$ were used for training, while the remaining $10\%$ were used as test set. Fig. \ref{fig:forward}a shows the comparison between the Poisson's ratios obtained by the FEA and the forward model $\mathcal{F}$ for both the train and test parts of \textit{Set 1}. Predictions of the network are very close to simulation results and a good generalization can be observed with data from train and test sets showing similar distributions ($R^2=0.99984$).
    
	\subsection{Set 2}
 
    \textit{Set 2} consists of 25,000 data points, of which again $90\%$ were used for training and $10\%$ for testing.
	Figure \ref{fig:forward}b shows the comparison between the Poisson's ratio calculated through FEA and the value predicted by the network. As for the optimal model, the predicted value should match the one from simulation, so ideally the plot would yield a line with unit slope. Similar to \textit{Set 1} the model generalizes properly, with test set performance equivalent to the training set ($R^2=0.99976$). 
    \\[1em]
    The obtained models show an extremely good prediction of the mechanical properties even with a relatively small number of samples. However, for both datasets multiple unit cells
    can yield the same Poisson's ratio. This causes the inverse problem to be ill-posed, as multiple solutions exist for a single input.

 	\begin{figure}[t]
		\centering
		\includegraphics[width=\textwidth]{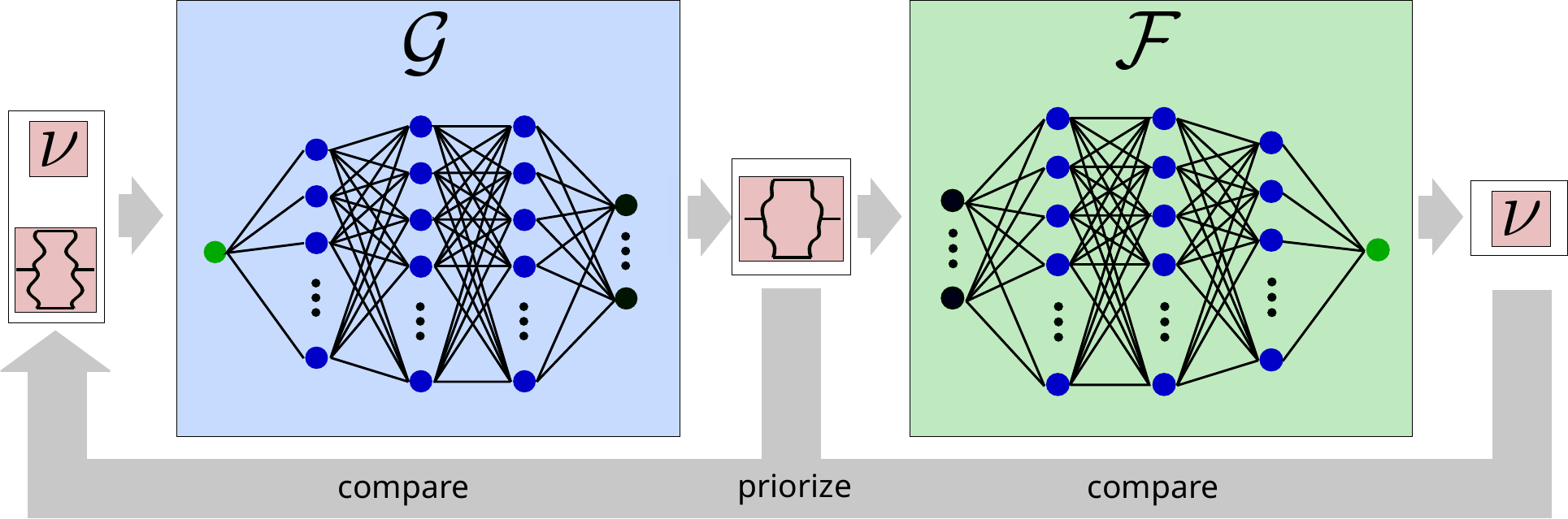}  
		\caption{Schematic of the inverse design system. The generative model $\mathcal{G}$ receives a target Poisson's ratio and guide curve. Based on these it generates a unit cells defined by a B-spline. This B-spline is then fed to the forward model $\mathcal{F}$ which estimates the Poisson's ratio of the generated structure so it can be compared to the target one for training. }
		\label{fig:inv_system}
	\end{figure} 
	
	\section{ML implementation for inverse (Property-Structure) problem}

    Solving the inverse problem is far more challenging than its forward equivalent, as it is not possible to simply reverse $\mathcal{F}$ and train this network on property/structure pairs. Instead it needs to be modified to address the following number of challenges, yielding the complex architecture consisting of interconnected inverse $\mathcal{G}$ (see Fig. \ref{fig:fwd_inv_model}b) and forward $\mathcal{F}$ (see Fig. \ref{fig:fwd_inv_model}a) models. 
    
    \textit{1)} Measures used to minimize the distance between a generated and a desired structure do not necessarily reflect similar material properties. A common approach to solve this is to utilize a pretrained surrogate model $\mathcal{F}$ \cite{kumar_inverse_2020}. Via $\mathcal{F}$ it is possible to backpropagate a property based loss to the generator network $\mathcal{G}$. Combining $\mathcal{F}$ and $\mathcal{G}$ in this manner yields the system shown in Fig. \ref{fig:inv_system}.
    
    \textit{2)} As mentioned in the previous section,  multiple structures can have the same properties, but a neural network can only produce one solution for a given input. Therefore, to allow multiple solutions additional inputs have to be added. This is often done in form of noise, where each component of the noise has to be connected to the generated structure, or $\mathcal{G}$ will simply learn to ignore it. This connection needs to be learned as well, which requires a lot of additional effort (for example by training a Generative Adversarial Network (GAN) \cite{goodfellow_generative_2020}). Instead of adding noise and learning a connection, here the additional input is given in form of a target B-spline that should be resembled by the generated curve.
    
    \textit{3)} As neural networks are usually not able to make accurate predictions outside of the range of their training data, it is necessary to limit generated structures to a sensible range when fitting the inverse model $\mathcal{G}$. Here an additional term was added to the loss function, that penalizes solutions which differ from a straight line between start and endpoint.   

    Each of these challenges is addressed by a distinct term of the following loss function:
	\begin{align}
        \begin{split}
            \min_{\mathcal{G}} \mathcal{O}(\mathcal{F}, \mathcal{G}) =  & \underbrace{\frac{1}{n} \sum_{i=1}^{n} (\mathcal{F}(\mathcal{G}(\textbf{x}_i,\textbf{y}_i))-\textbf{y}_i)^2}_{1)} + \alpha \underbrace{ \frac{1}{n} \sum_{i=1}^{n}  (\mathcal{G}(\textbf{x}_i,\textbf{y}_i)-\textbf{x}_i)^2}_{2)} \\
            & + \underbrace{\frac{1}{n} \sum_{i=1}^{n} \sum_{j=1}^{n_c} \mathbbm{1}_{d_j>K} ~ d_j}_{3)} ~,
        \end{split}
        \label{eq:inv}
    \end{align}
    where $n$ is the number of samples, $n_c$ the number of control points, $d_j = (\mathcal{G}(\textbf{x}_i)_j - \frac{j-1}{n_c-1}c_{n_c}  )^2$ the deviation of the B-spline from a straight line and $K$ a threshold for penalizing this deviation. 
    \\[1em]
    In Eq. \ref{eq:inv} the parameter $\alpha$ controls the trade-off between matching the target properties or matching the guide curve (see left side of Fig. \ref{fig:inv_system}). For training the inverse model $\mathcal{G}$, both target properties and guide curve were sampled from random uniform distributions. Note, that the guide curve is an arbitrary curve that might be not represented in the dataset. The network shown in Fig. \ref{fig:fwd_inv_model}b was trained separately on \textit{Set 1} and \textit{Set 2}, each time combined with the respective forward model. For \textit{Set 1} the end points of the guide curves as well as the ones of the generated B-splines were fixed. Similar to training the forward model, Adam was used as optimizer. 

    \begin{figure}[t]
        \centering
        \begin{minipage}[t]{0.49\textwidth}		
            ~ \textbf{(a)} \\
            \includegraphics[width=\textwidth]{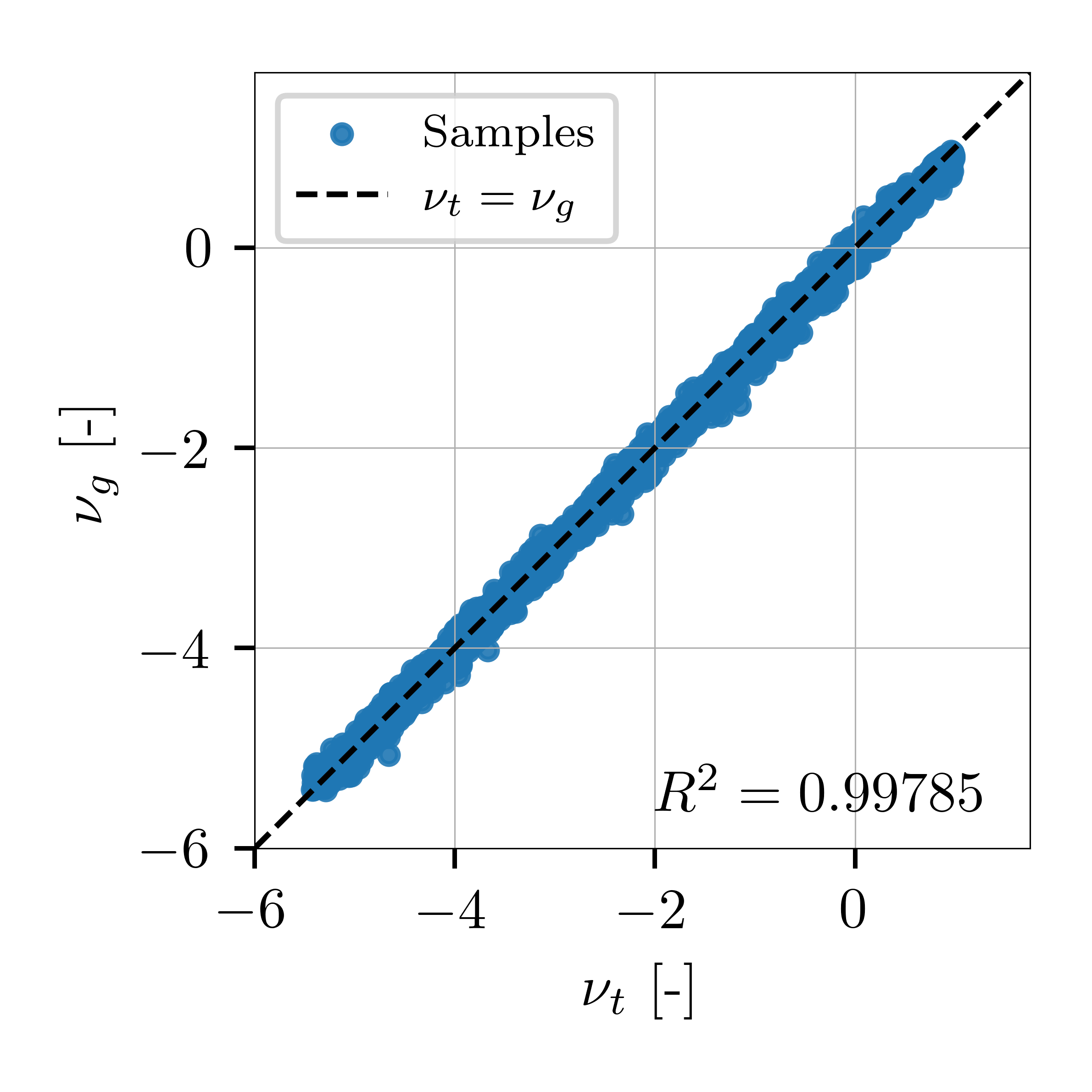}
        \end{minipage}   
        \begin{minipage}[t]{0.49\textwidth}		
            ~ \textbf{(b)} \\  
            \includegraphics[width=\textwidth]{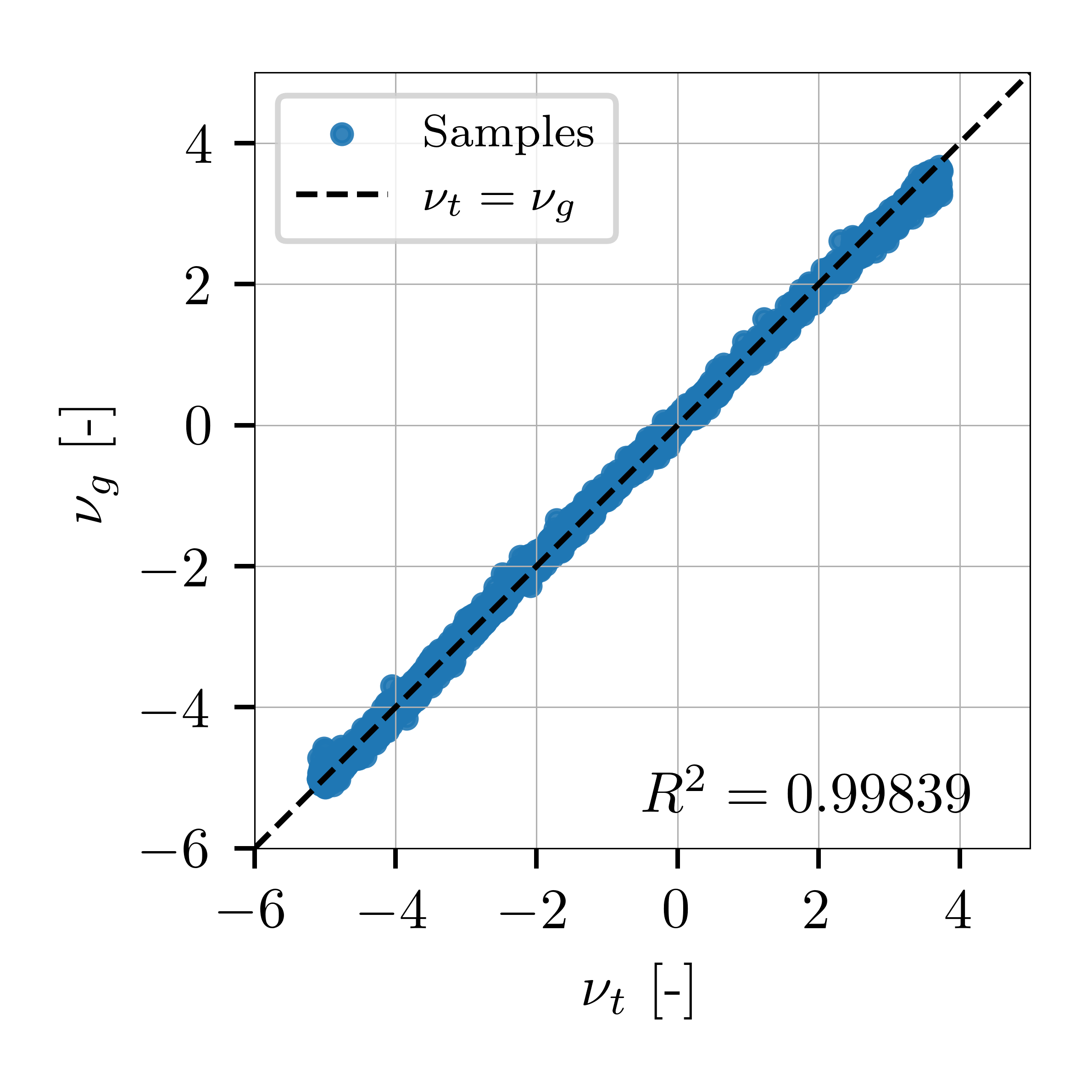}
        \end{minipage}   
        \caption{Comparison between the target Poisson's ratios ($\nu_{t}$) and the Poisson's ratios of the corresponding structures by $\mathcal{G}$ ($\nu_{g}$) for training and testing data from \textit{Set 1} \textbf{(a)} and \textit{Set 2} \textbf{(b)}.}
        \label{fig:inverse}
    \end{figure}	

    Fig. \ref{fig:inverse} shows the comparison between target and resulting mechanical properties (Poisson's ratio) for randomly picked guide curves. The target properties were chosen randomly, while the resulting Poisson's ratios of the corresponding generated curves were obtained through FEA. A good fitting ($R^2=0.99785$ for \textit{Set 1} and $R^2=0.99839$ for \textit{Set 2}) was obtained, although $\mathcal{G}$ was not directly trained on FEA data, but $\mathcal{F}$ instead. As expected, the inverse model trained on \textit{Set 2} can design unit cells with a greater variety of properties than the model trained on \textit{Set 1}.

\begin{figure}
        \centering
        \begin{minipage}[t]{0.49\textwidth}		
            \textbf{(a)} \\[0.5em]
            \includegraphics[width=\textwidth]{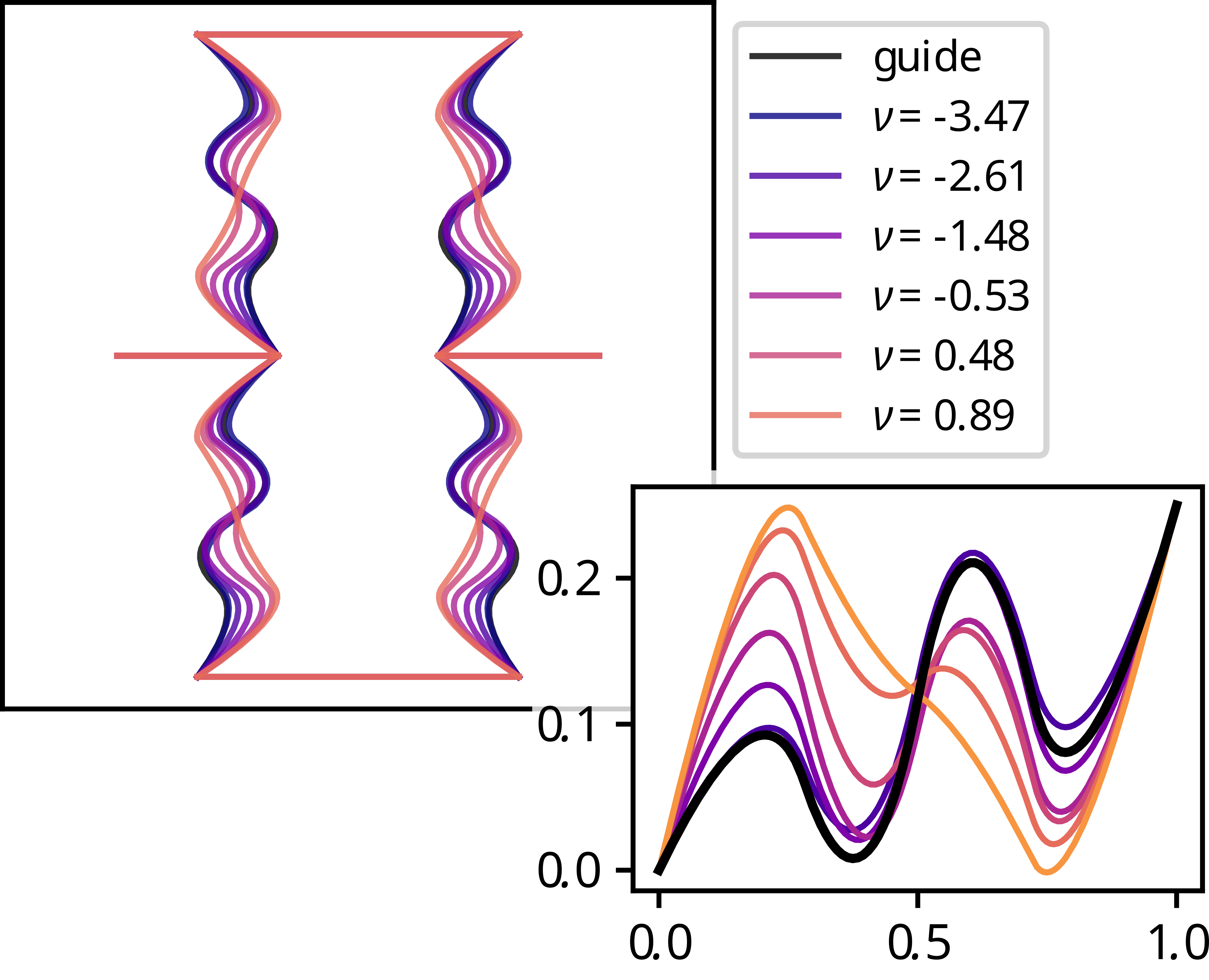}
            \vspace{0.5em}
        \end{minipage}   
        \begin{minipage}[t]{0.49\textwidth}		
            \textbf{(b)} \\[0.5em]  
            \includegraphics[width=\textwidth]{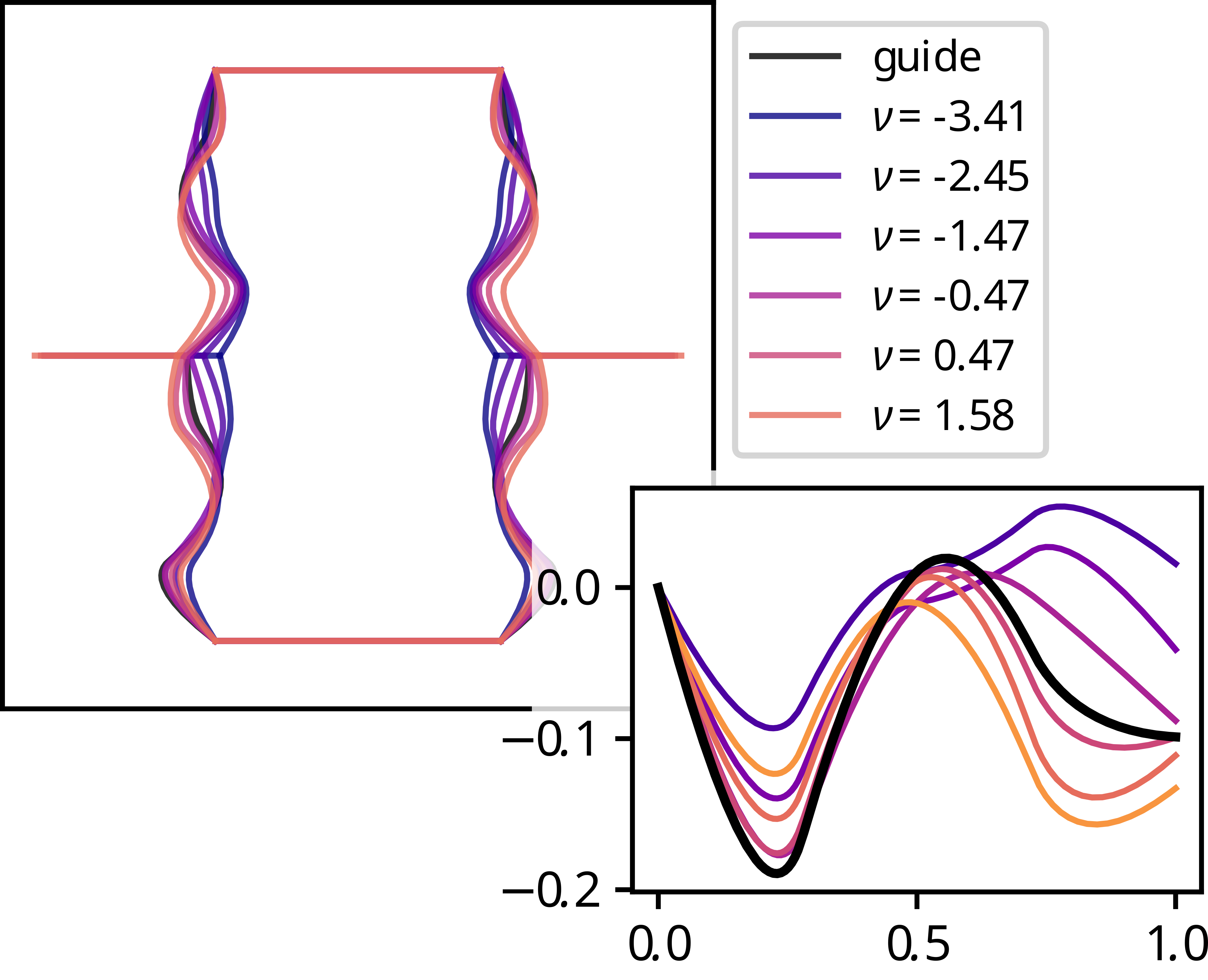}
            \vspace{0.5em}
        \end{minipage}  
        
        \begin{minipage}[t]{0.49\textwidth}		
            \textbf{(c)} \\[0.5em]
            \includegraphics[width=\textwidth]{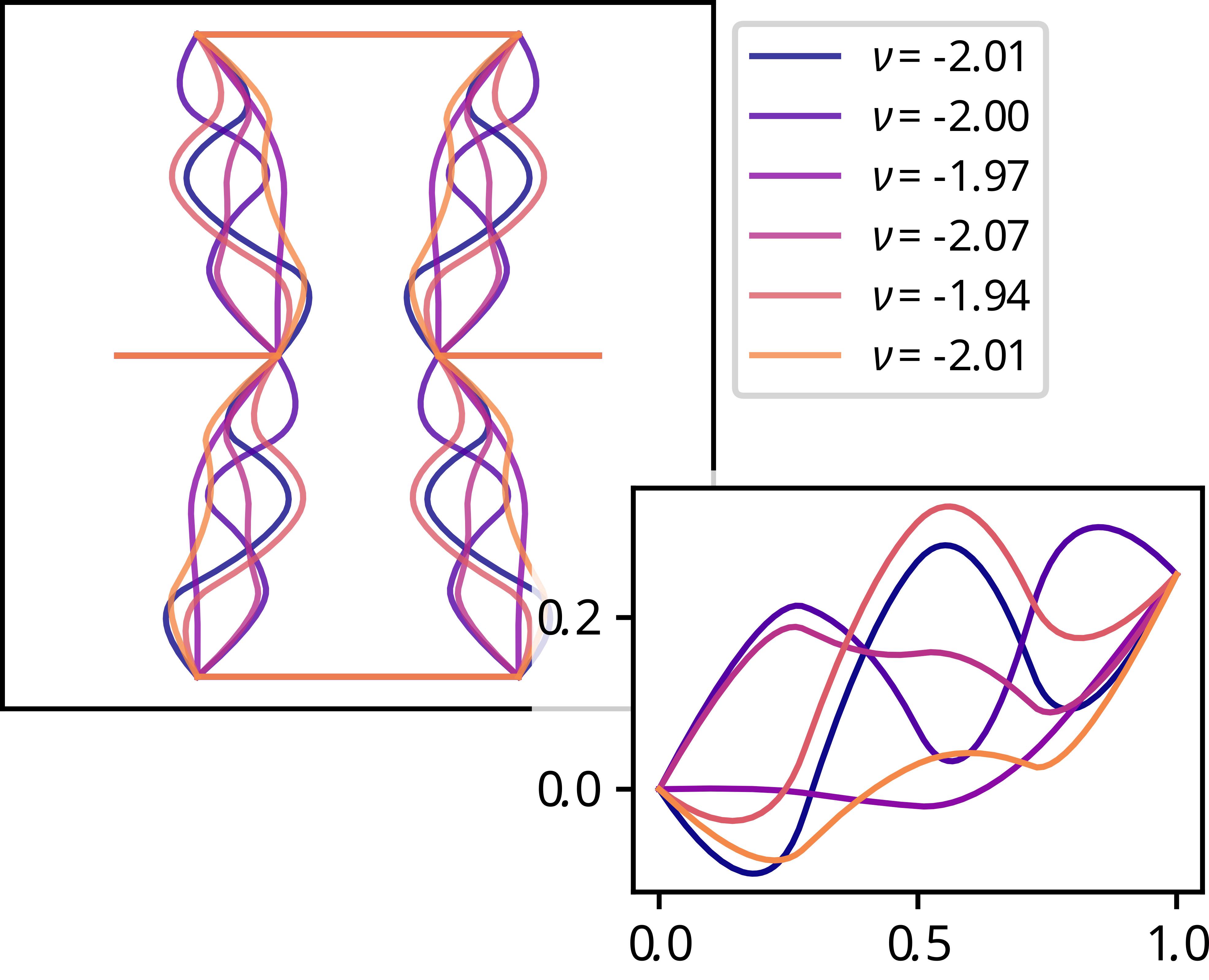}
        \end{minipage}   
        \begin{minipage}[t]{0.49\textwidth}		
            \textbf{(d)} \\[0.5em]  
            \includegraphics[width=\textwidth]{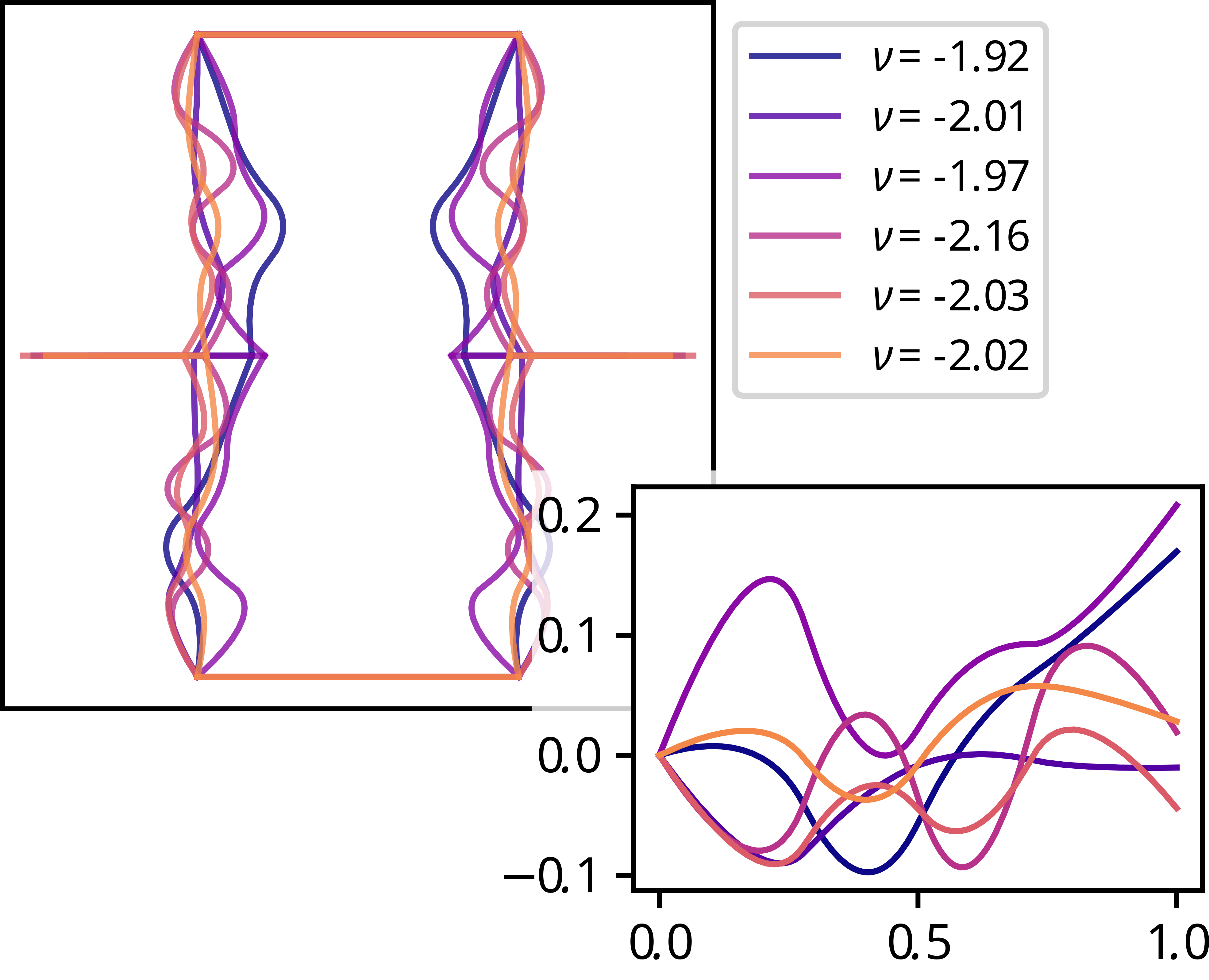}
        \end{minipage}    
    
        \caption{Generated unit cells and corresponding B-splines for variation of target Poisson's ratio with the same guide curve for \textit{Set 1} \textbf{(a)}  and \textit{Set 2}  \textbf{(b)}. Similarly, the generated unit cells and corresponding B-splines for variation of the guide curve while keeping the target Poisson's ratio the same for \textit{Set 1} \textbf{(c)}  and \textit{Set 2} \textbf{(d)}.}  
        \label{fig:set-curves}   
    \end{figure}

    Since $\mathcal{G}$ has two inputs (target Poisson's ratio and guide curve), we can demonstrate the performance of the inverse network in two different ways. Fig. \ref{fig:set-curves}a-b show the variety of curves generated for the same guide curve with different target Poisson's ratio. It can be observed that a continuous change in Poisson's ratio usually leads to quite smooth transformations in the resulting curves. However, for some values of Poisson's ratio, one cannot maintain the similarity with the guide curve. This is in part due to the additional constraint on the "width" of the curve (see last term in Eq. \ref{eq:inv}) 
    
    Similarly Fig. \ref{fig:set-curves}c-d show the variety of the generated curves with the same properties, but based on the different guide curves. It can be observed, that the variety of the admissible generated curves with the same Poisson's ratio is immense. Note, that the model trained on \textit{Set 2} shows a wider variety of curves and Poisson's ratios as compared to the model trained on \textit{Set 1}. This is expected behavior since geometrically \textit{Set 1} is the subset of \textit{Set 2}, which includes unit cells based on hexagonal elements ($\theta>0$) besides of reentrant ones. 

    \section{Experiments}
\begin{figure}[ht]
        \centering
        \begin{minipage}[t]{0.49\textwidth}		
            \textbf{(a)} \\[0.7em]    
            \begin{minipage}[t]{\textwidth}		
                \centering
                \includegraphics[width=0.65\textwidth]{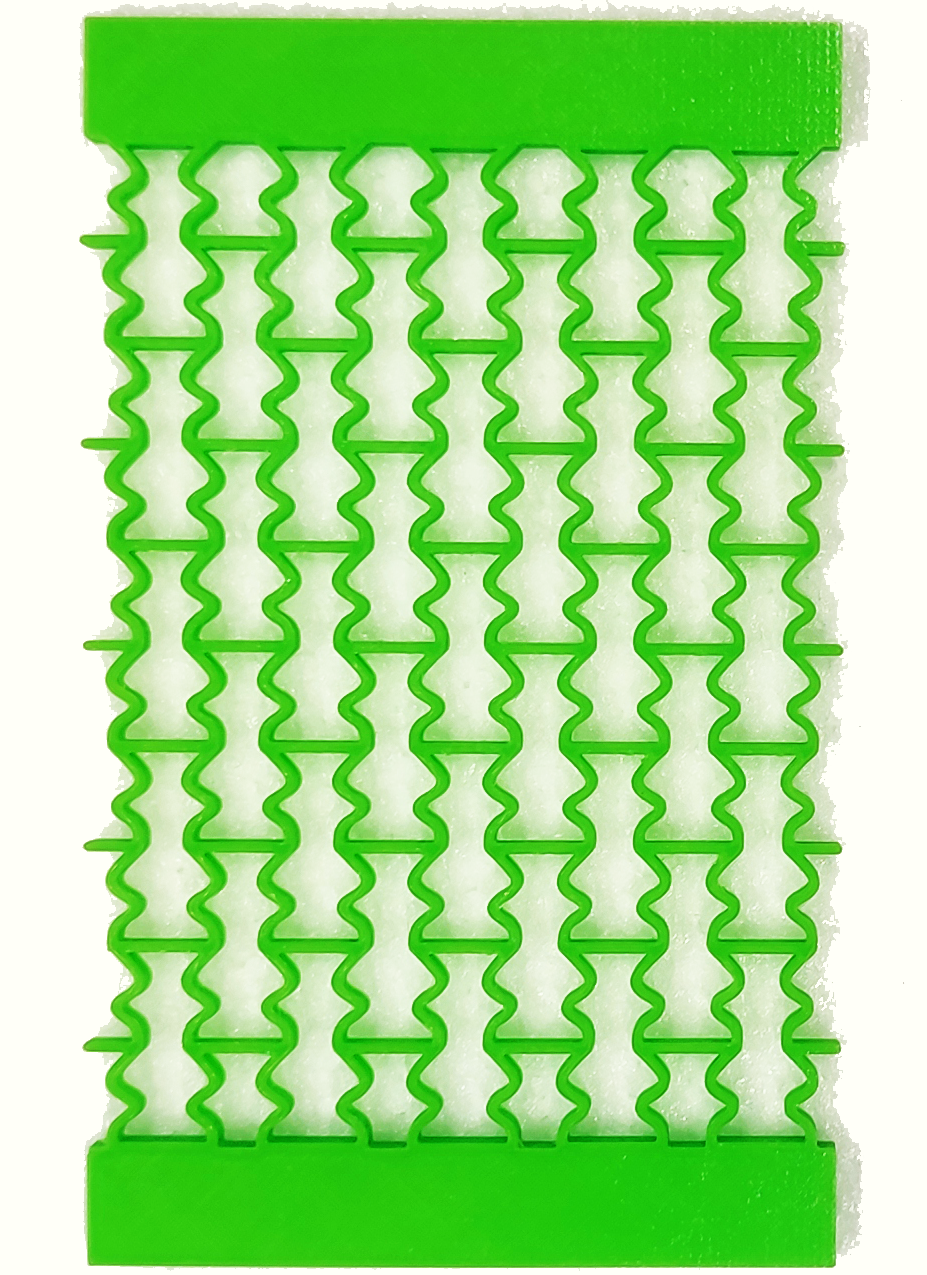}
            \end{minipage}   
        \end{minipage}       
        \begin{minipage}[t]{0.49\textwidth}		
            ~ \textbf{(b)} \\  
            \includegraphics[width=\textwidth]{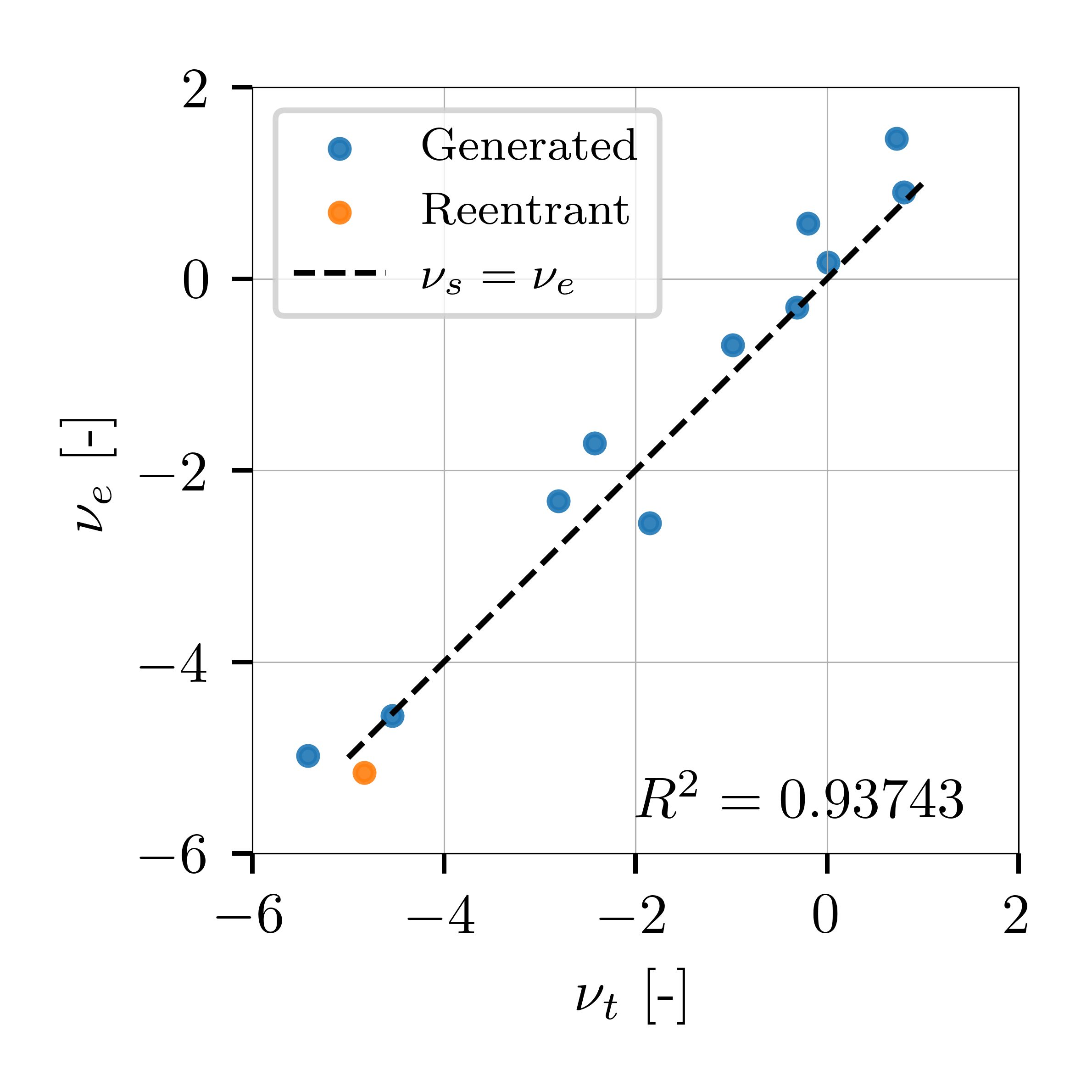}
        \end{minipage}   
        \caption{ An exemplary test specimen \textbf{(a)}. Comparison between the target Poisson’s ratios ($\nu_t$) and the Poisson’s ratios derived from the experiments ($\nu_e$) \textbf{(b)}.}
        \label{fig:experiment}
    
    \end{figure}

    To validate the finite element simulation, mechanical tests were performed on ten different geometries. While nine out of these ten specimens were based on unit cells generated by the inverse model $\mathcal{G}$ for \textit{Set 1}, the tenth sample was based on the corresponding reentrant unit cell with straight beams for reference. These specimens (see Fig. \ref{fig:experiment}a) consisted of 5x5 unit cells ($25\text{mm} \times 18.75\text{mm} \times 5\text{mm}$ each) were additively manufactured from $\text{Ninjaflex}^{\text{\textregistered}}$ TPU filament by a Creality CR-10 3D-Printer. The tensile behavior of the samples was tested on a Zwick\&Roell 10kN universal testing machine. Since the purpose of these tests was to verify how well the real specimens fit the target Poisson's ratio, specimens included structures with both auxetic and non-auxetic behavior. To minimize viscoelastic effects and assure conditions as close to the simulations as possible, samples were subjected to a tensile load with a ratio of $20~\text{mm/min}$. The resulting Poisson's ratio was determined as quotient of lateral and applied strain  at $1\%$ total strain. The measurements necessary for this calculation were taken by analysing captured videos of the experiments in the software Fiji \cite{schindelin_fiji_2012}. 

    The test results shown in Fig. \ref{fig:experiment}b display a clear correlation between target and exhibited behavior. This means that despite being only trained on synthetic data, our model is able to make reliable predictions in a real world setting.
    
    \section{Conclusion}
    
    In this work, we introduced a class of reentrant-hexagonal metamaterials based on curved Bezier beams together with a deep learning framework to generate unit cells with specific mechanical properties. Using Bezier curves instead of straight beams allowed us to obtain a large variety of different properties, while keeping the dimensions of the corresponding unit cells constant. At the same time, it yields a large number of possible unit cells to fit given properties. We used a guide curve instead of noise to create a one-on-one mapping for this inverse problem. This approach enables us not only to generate several unit cells with the same properties, but also to express preference for a specific shape.
     
    As the proposed forward neural network model can reliably predict mechanical properties of metamaterials ($R^2 > 0.999$), the corresponding inverse model was able to accurately generate unit cells fitting specific properties ($R^2>0.997$). Experiments showed that this ability extends to a real world setting, even though both models being trained on simulations.
    Furthermore, both models are computationally efficient, allowing us to generate a large number of unit cells in a short amount of time (around $\text{20,000}$ unit cells per second). 

    \section{Acknowledgement}
    Funded by the Deutsche Forschungsgemeinschaft (DFG, German Research Foundation) under Germany’s Excellence Strategy – EXC-2193/1 – 390951807. The authors acknowledge support by the state of Baden-Württemberg through bwHPC and the German Research Foundation (DFG) through grant no INST 39/963-1 FUGG (bwForCluster NEMO).
    	
    \bibliographystyle{plainnat}
    \bibliography{lib2}
	
\end{document}